\documentclass[12pt,aps,amssymb,noshowpacs,superscriptaddress,nofootinbib]{revtex4-1}
\usepackage{graphicx}
\usepackage{enumerate}

\begin{document}
\title{Hamiltonian Analysis of $R + T^2$ Action}
\author{Jian Yang}
\email{yjccnuphy@yahoo.com.cn}
\affiliation{School of Science,
Beijing University of Posts and Telecommunications,\\
Beijing 100876, China.}
\author{Kinjal Banerjee}
\email{kinjalb@gmail.com}
\author{Yongge Ma\footnote{Corresponding author}}
\email{mayg@bnu.edu.cn}
\affiliation{Department of Physics, Beijing Normal University, \\Beijing 100875, China.}

\date{\today}
\begin{abstract}
We study the gravitational action which is a linear combination of the Hilbert-Palatini term and a term quadratic in torsion and possessing local
Poincare invariance. Although this action yields the same equations of motion as General Relativity, the detailed Hamiltonian analysis without gauge
fixing reveals some new points never shown in the Hilbert-Palatini formalism. These include that an additional term containing torsion appears in the
spatial diffeomorphism constraint and that the primary second-class constraints have to be imposed in a manner different from that in the
Hilbert-Palatini case. These results may provide valuable lessons for further study of Hamiltonian systems with torsion.
\end{abstract}

\maketitle

\def\ba{\begin{eqnarray}}
\def\ea{\end{eqnarray}}
\def\del{\partial}

\section{Introduction}

Among various attempts to look for a quantum gravity theory, gauge theories of gravity are very attractive since the idea of gauge invariance has already
been successful in the foundation of other fundamental interactions. Local gauge invariance is a key concept in Yang-Mills theory. Together with
Poincare symmetry, it lays the foundation of standard model in particle physics. Localization of Poincare symmetry leads to Poincare Gauge Theory(PGT) of
gravity, which contains general relativity as a special case. In this theory, besides energy-momentum, the spin of matter fields is also introduced
to take effect on gravitational dynamics. It turns out that from the perspective of PGT, in general, gravity is not only represented as curvature but
also as torsion of space-time. (See \cite{hehl} and references therein for a comprehensive account of torsion in gravity). A large number of actions which
satisfy local Poincare symmetry have been analyzed by various researchers (\cite{PGT} provides a comprehensive review and bibliography of the
progress made in PGT).

Hamiltonian formalism is an inherent framework to study the dynamics of a physical theory. Although Hamiltonian analysis is performed for a
large number of models in PGT, the results are at a formal level without explicit expressions of the additional required second-class constraints.
From the point of view of canonical quantization, it is essential to have a well-defined consistent Hamiltonian theory at the classical level. Such
an ingredient is missing if we want to incorporate torsion into candidate quantum gravity models. However, it has been shown \cite{shyam,mercuri} that
the connection dynamics of general relativity can be derived from an action which contains, apart from the standard Hilbert-Palatini term, a total
derivative involving torsion known as the Nieh-Yan term.

In this paper, we are interested in an action which contains two terms, a standard Hilbert-Palatini term and a $T^2$ term which involves a
product of two torsion tensors in a particular way. There are a couple of reasons for choosing this particular $T^2$ term. This term has not been
analyzed in the literature of PGT. Also this term is actually the difference between the Holst term \cite{holst,sa} and Nieh-Yan term \cite{niehyan}.
(For details see \cite{shyam,kinjal}.) Although it turns out that the only solutions of this action are torsion free, our analysis is still very
interesting as we manage to explicitly determine all the second-class constraints unlike other attempts in PGT. There are several subtle and important
points in our Hamiltonian analysis which may provide valuable lessons for further studies of Hamiltonian systems with torsion. In the presence of
torsion, an additional term containing torsion appears in the spatial diffeomorphism constraint. Moreover the primary second-class constraints
appearing in the first-order formalism have to be imposed in a manner different from that in the Hilbert-Palatini theory without torsion.

The paper is organized as follows. We first perform the Lagrangian analysis based on action principle in Section \ref{sec2}. Then in Section
\ref{sec3}, we perform the Hamiltonian analysis of the theory in a way which is very different from that in the Hilbert-Palatini theory as well as
the one carried out in \cite{shyam} for a related action. Our approach is more closely related to the approach taken in PGT \cite{PGT}. This is the main
section of the paper and our analysis indicates several new and important properties which may be useful for further analysis of any other action with
torsion. Finally we end with a conclusion in Section \ref{sec4}.

We will restrict ourselves to 4 dimensions. The Greek letters {$\mu, \nu \dots$} refer to space-time indices while the uppercase Latin letters
{$I,J \dots$} refer to the internal $SO(3,1)$ indices. Our spacetime metric signature is $(-+++)$. Later when we do the $3+1$ decomposition of spacetime, we will use the lowercase
Latin letters from the beginning of the alphabet {$a,b,\dots$} to represent the spatial indices.

\section{Action and Lagrangian Analysis}\label{sec2}

The action of gravity which we consider in this paper reads
\begin{eqnarray}
&&S = S_{HP} + \alpha S_T  \hspace{6em} \label{action}\\
\mbox{where}\hspace{2em}&& \nonumber \\
 &&S_{HP} =\int {\mbox d}^4 x ~eR= \int {\mbox d}^4 x e e^{\mu}_I e^{\nu}_J R_{\mu \nu}^{~~ IJ}(\omega_\mu^{~IJ}) ~~;~~
S_T  = \frac{1}{8}\int {\mbox d}^4 x \epsilon^{\mu \nu \rho \sigma} T^{I}_{~ \mu \nu} T_{I \rho \sigma } . \nonumber
\end{eqnarray}
Here the coupling parameter $\alpha$ is a non-zero real number, $e^{\mu}_I$ is the tetrad, $e$ denotes the absolute value of the determinant of
the co-tetrad, $\omega_\mu^{~IJ}$ is the spacetime spin-connection which is not torsion-free, and $\epsilon^{\mu \nu \rho \sigma}$ denotes the
4-dimensional Levi-Civita tensor density. Further
\begin{eqnarray}
R_{\mu \nu}^{~~ IJ} &=& \del_{[\mu} \omega_{\nu]}^{~IJ} + \omega_{[\mu}^{~IK} \omega_{\nu] K}^{~~~J} , \\
T^{I}_{~ \mu \nu}  &=& \del_{[\mu} e_{\nu]}^{I} +  \omega_{[\mu~|J|}^{~~I}e_{\nu]}^{J}
\end{eqnarray}
are the definitions for curvature and torsion respectively
\footnote{Our conventions of symmetrization and antisymmetrization
are $A^{(ab)}:=A^{ab} + A^{ba}$ and $A^{[ab]}:=A^{ab} - A^{ba}$
respectively}. This action is invariant under local Poincare transformations (see Appendix A). Moreover, it is easy to show that $S_T$  can be rewritten as
\begin{eqnarray}
S_T &=& \int {\mbox d}^4 x \left[-\frac{1}{8} e e^{[\mu}_I e^{\nu]}_J R_{\mu \nu~ KL}~ \epsilon^{IJKL} +
\del_\mu\left( \frac{1}{4} \epsilon^{\mu \nu \rho \sigma} e_{\nu}^I D_{[\rho}e_{\sigma] I} \right)  \right] \nonumber \\
&=& \int - \mathcal{L}_H + \mbox{total derivative} \nonumber
\end{eqnarray}
where $\mathcal{L}_H$ is known as the Holst term for generalized
Palatini action of gravity. It is well known that, in the absence of
matter, adding the Holst term to the Hilbert-Palatini action does
not modify the Einstein's equations of motion \cite{holst,sa}. So,
although the additional term in our action is not a total
derivative, it is related by a total derivative, known as the
Nieh-Yan term \cite{niehyan}, to a term which does not modify the
equations of motion. We now show that the action  (\ref{action})
also yields Einstein's equations as expected, and thus the coupling
parameter $\alpha$ is free at classical level.

We will be working in the first-order formalism and hence both the
co-tetrad $e_{\mu}^I$ and the spin connection $\omega_{\mu}^{~IJ}$ are treated as independent fields. Also our covariant derivative $D_\mu$ acts in
the following way:
\begin{eqnarray}
D_\mu e_\nu^I:= \partial_\mu e_\nu^I +  \omega_{\mu ~ J}^{~I} e_\nu^J . \nonumber
\end{eqnarray}
Then the variation of the action (\ref{action}) with respect to the basic variables give:
\begin{eqnarray}
\delta(S_{HP}) &=& \int {\mbox d}^4 x\bigg[ \bigg( e e^{\alpha}_K e^{\mu}_I e^{\nu}_J R_{\mu \nu}^{~~ IJ} -
2 e  e^{\alpha}_I  e^{\mu}_K e^{\nu}_J R_{\mu \nu}^{~~ IJ}\bigg)\delta e_{\alpha}^K \nonumber\\
&&\hspace{6em} - \bigg(D_{\mu}\left[\frac{1}{2} \epsilon^{\mu \nu \rho \sigma} \epsilon_{IJKL} e_{\rho}^K e_{\sigma}^L \right]\bigg)
\delta\omega_{\nu}^{~IJ} \bigg] , \label{deltaSHP} \\
\alpha \delta(S_T)  &=& \int {\mbox d}^4 x \bigg[ \frac{\alpha}{2}\bigg(D_{\beta}
\left[ \epsilon^{\alpha \beta \gamma \delta} D_{[\gamma}e_{\delta]K}\right] \bigg)
\delta e_{\alpha}^K + \bigg( \frac{\alpha}{2} \epsilon^{\mu \nu \rho \sigma} e_{\nu J} D_{[\rho}e_{\sigma]I}\bigg)\delta\omega_{\mu}^{~IJ} \bigg]
\nonumber\\\label{deltaST}
\end{eqnarray}
where we have dropped the total derivatives. Thus it is easy to determine the equations of motion of the full action.
The variation of the spin connection $\omega_{\mu}^{~IJ}$ yields
\begin{eqnarray}
\frac{1}{2}\epsilon^{\mu \nu \rho \sigma}  e_{\nu}^K D_{[\rho}e_{\sigma]}^L
\left[\frac{\alpha}{2} (\eta_{JK} \eta_{IL} -\eta_{IK} \eta_{JL}) - \epsilon_{IJKL}\right] = 0 \label{lageom1}
\end{eqnarray}
Direct calculation shows that this implies $\epsilon^{\mu \nu \rho \sigma}  e_{\nu}^K D_{[\rho}e_{\sigma]}^L =0$, and hence similar to that in \cite{peldan}, the only
solutions are $ D_{[\rho}e_{\sigma]}^L = 0$.
\footnote{Note that, although the covariant derivative in \cite{peldan} is torsion free and different from our covariant derivative, this result
still holds in the case there is an antisymmetrization in the spacetime indices.} So, in the absence of matter the only solutions are the torsion-free
solutions.

On the other hand, the variation of the co-tetrad $e_{\alpha}^K$ leads to
\begin{eqnarray}
e e^{\alpha}_K e^{\mu}_I e^{\nu}_J R_{\mu \nu}^{~~ IJ} - 2 e  e^{\alpha}_I  e^{\mu}_K e^{\nu}_J R_{\mu \nu}^{~~ IJ} +
\frac{\alpha}{2} \left(D_{\beta}[\epsilon^{\alpha \beta \gamma \delta} D_{[\gamma}e_{\delta]K}]\right) = 0 \label{lageom2}
\end{eqnarray}
We then get back the standard Einstein's equation for co-tetrad from (\ref{lageom2}) after solving for spin connection by Eq.(\ref{lageom1}).
Hence, although the action with which we started contains the $T^2$ term, the equations of motion are as same as those of the
Hilbert-Palatini action. In the next section we shall perform the Hamiltonian analysis of this action.

\section{Hamiltonian analysis} \label{sec3}

In the Hamiltonian formulation of Hilbert-Palatini theory the basic variables are the $SO(3,1)$ spin connection $\omega_a^{~IJ}$ and its
conjugate momentum. It is well known that this formulation contains second-class constraints. Since our action contains the other term
which explicitly depends on torsion, we expect that there will be the other pair of conjugate variables and the second-class constraints
will be somehow different from the Hilbert-Palatini case.

To seek a complete Hamiltonian analysis, we perform the $3+1$ decomposition of our fields without breaking the internal $SO(3,1)$ symmetry and
also without fixing any gauge. To identify our configuration and momentum variables for performing Hamiltonian analysis, we can rewrite
$S_{HP}$ and $S_T$ as:
\begin{eqnarray}
S_{HP} &=& \int {\mbox d}^4 x \bigg[  e e^t_{[I}e^a_{J]} \left(\del_{t} \omega_a^{~IJ}\right) + e e^t_{[I}e^a_{J]} \left(-\del_{a} \omega_t^{~IJ}
+  \omega_{[t}^{~IK} \omega_{a]}^{~KJ}\right)  + \frac{1}{2} e e^{a}_{[I} e^{b}_{J]} R_{a b}^{~~ IJ} \bigg], \nonumber\\ \\
\alpha S_{T} &=& \alpha\int {\mbox d}^4 x \bigg[\epsilon^{abc} D_{b}e_{c}^I \left(\del_{t} e_a^I\right)
+ \epsilon^{abc} D_{b}e_{c}^I \left(-\del_{a} e_t^I + \omega_{[t}^{~~IJ}e_{a]J} \right)\bigg]
\end{eqnarray}
So we can identify the momenta with respect to $\omega_a^{~IJ}$ and $e_a^I$ as
\begin{eqnarray}
\Pi^a_{IJ} :=  e e^t_{[I}e^a_{J]} ~~~~;~~~~ \Pi^a_I := \alpha \epsilon^{abc} D_{b}e_{c I} \label{momentumdef1}
\end{eqnarray}
where $\epsilon^{abc}$ denotes the 3-dimensional Levi-Civita tensor density.
We further parametrize the tetrad and the co-tetrad fields as \cite{peldan}:
\begin{eqnarray}
e_{t I} = N N_I + N^a V_{a I} ~~~~ &;& ~~~~ e^{t I} = - \frac{N^I}{N} ,\nonumber \\
e_{a I} = V_{a I} ~~~~ &;& ~~~~ e^{a I} = V^{a I} + \frac{N^a N^I}{N} ,\label{newparameters1}\\
\mbox{with}\hspace{4em}
N^I V_{a I} = 0 ~~~~ &;& ~~~~  N^I N_I =-1  \label{parameterconstraints1}
\end{eqnarray}
What we have done is that we have reparametrized the 16 degrees of freedom of $e_{\mu I}$ into 20 fields given by (\ref{newparameters1}) subject to
the 4 constraints (\ref{parameterconstraints1}). From these definitions, the following identities also hold:
\begin{eqnarray}
&&V^{a I} V_{b I} = \delta^a_b ~~~~ ; ~~~~  V^{a I}N_I = 0  ~~~~ ; ~~~~ N_a := V_{a I} V_{b}^I N^b ,\nonumber\\
&& V^{aI} V_{a}^{~J} = \eta^{IJ} + N^I N^J   \label{parameterconstraints2}
\end{eqnarray}
In terms of these fields the metric takes the standard form
\begin{eqnarray}
g_{\mu \nu} = \left(\begin{array}{cc}
              -N^2 + N^a N_a  &    N_a    \\
               N_a            &  V_{a I} V_b^I
           \end{array}\right)                   \nonumber
\end{eqnarray}
It is easy to see that
\begin{eqnarray}
g &:=& \mbox{det} (g_{\mu \nu}) = - N^2 \mbox{det}( V_{a I} V_b^I ),\nonumber \\
e &:=& \mbox{det} (e_{\mu I}) = N \sqrt{\mbox{det}( V_{a I} V_b^I) }
= N \sqrt{\mbox{det}( q_{ab})} \equiv N \sqrt{q},\nonumber
\end{eqnarray}
where $q_{ab} := V_{a I} V_b^I$ is the induced 3-metric on spacelike hypersurfaces $\Sigma_t$ in the $3+1$ decomposition of spacetime.

Using the definitions given above we can also prove the following two identities
\begin{eqnarray}
- e e^{a}_{[I} e^{b}_{J]} &=& \frac{N^2}{e} \Pi^{[a}_{IK} \Pi^{b]}_{JL} \eta^{KL} + N^{[a}\Pi^{b]}_{IJ} , \label{identity1} \\
V^a_I &=& - \frac{1}{\sqrt{q}}\Pi^a_{~IJ} N^J \label{identity2}
\end{eqnarray}
Using these identities, we can rewrite the actions as
\begin{eqnarray}
S_{HP} &=& \int {\mbox d}^4 x \bigg[\Pi^{a}_{IJ} \del_t  \omega_a^{~IJ} -
\left( \frac{N^2}{2 e} \Pi^{[a}_{IK} \Pi^{b]}_{JL} \eta^{KL}  R_{a b}^{~~ IJ}  + \frac{1}{2}N^{[a}\Pi^{b]}_{IJ}  R_{a b}^{~~ IJ}
-\omega_t^{~IJ}D_a \Pi^{a}_{IJ}  \right) \bigg] ,\nonumber\\ \\
\alpha S_{T} &=& \int {\mbox d}^4 x \bigg[ \Pi^a_I \del_t V_a^I + \left( N N^I D_a \Pi^a_I +N^a V_a^I D_b \Pi^b_I +
\frac{1}{2}\omega_t^{~IJ} \Pi^a_{[I} V_{J]a} \right) \bigg]
\end{eqnarray}
where the total derivative terms have been neglected.
It is going to be clear that the torsion degrees of freedom are now encoded in $\Pi^a_I$. In particular, if $\Pi^a_I = 0$ the entire $S_T$
vanishes and we are left with the torsion-free solutions. Also the fact that $V_a^I$ is now a basic variable causes significant changes in the
subsequent constraint analysis as compared to the Hilbert-Palatini case.

Now we are in a position to rewrite the total action in the ADM form
\begin{eqnarray}
S_{HP} + \alpha S_{T} &=& \int {\mbox d}^4 x \bigg[\Pi^{a}_{IJ} \del_t  \omega_a^{~IJ} + \Pi^a_I \del_t V_a^I - \left(N H + N^a H_a +
\omega_t^{~IJ} \mathcal{G}_{tIJ} \right) \bigg] \nonumber \\
\mbox{where} \hspace{4em}
H &=&   \frac{1}{\sqrt{q}} \Pi^{a}_{IK} \Pi^{b}_{JL} \eta^{KL}  R_{a b}^{~~ IJ}  -  N^I D_a \Pi^a_I ,\label{hamiltonian1}\\
H_a &=&  \Pi^{b}_{IJ}  R_{a b}^{~~ IJ} - V_a^I D_b \Pi^b_I , \label{diffeo1}\\
 \mathcal{G}_{tIJ} &=& - D_a \Pi^{a}_{IJ} - \frac{1}{2}\Pi^a_{[I} V_{J]a} \label{gauss1}
\end{eqnarray}
Subsequently we will drop the subscript $t$ from $\mathcal{G}_{tIJ}$ and denote it as $\mathcal{G}_{IJ}$.
The two pairs of conjugate variables are $\left( \Pi^a_{~IJ}, \omega_a^{~IJ} \right)$ and $\left( \Pi^a_I,V_a^I  \right)$.
The fundamental Poisson brackets are given by
\begin{eqnarray}
\left\{  \omega_a^{~IJ}(x),\Pi^{b}_{KL}(y) \right\} &=& \frac{1}{2} \delta_a^b \delta_K^{[I} \delta^{J]}_L \delta^3 (x-y) ,  \nonumber \\
\left\{  V_a^I(x),\Pi^b_J(y) \right\} &=& \delta_a^b \delta^I_J \delta^3 (x-y)
\end{eqnarray}

Now let us count the degrees of freedom. The pair $\left( \Pi^a_{~IJ}, \omega_a^{~IJ} \right)$ have 36 degrees of freedom while the pair
$\left( \Pi^a_I,V_a^I  \right)$ have 24. The total number of degrees of freedom are 60. In the Lagrangian analysis we have seen that the only
solutions are the standard torsion-free solutions. Hence, if the Lagrangian and Hamiltonian formulations are to be equivalent, the constraints
present in the theory should remove 56 degrees of freedom leaving only 4 per point.
At this stage we have the following constraints:
\begin{itemize}
\item Since there is no momentum corresponding to $\omega_t^{~IJ}$, we have to impose 6 primary constraints $\Pi^t_{IJ} \approx 0$.
This leads to 6 secondary constraints $\mathcal{G}_{IJ} \approx 0$.
\item Also there is no momentum corresponding to $e_t^I$. We have to impose 4 constraints $\Pi^t_{I} \approx 0$ which lead to 3 secondary
constraints $H_a \approx 0$ and 1 more secondary constraint $H \approx 0$.
\item From Eqs (\ref{momentumdef1}), we can get two other sets of primary constraints
\begin{eqnarray}
C^a_I &:=& \Pi^a_I - \alpha \epsilon^{abc} D_{b}V_{c I} \approx 0 , \label{newconstraint1}\\
\Phi^a_{IJ} &:=& \Pi^a_{IJ} -\frac{1}{2} \epsilon^{abc} \epsilon_{IJKL} V_b^K V_c^L \approx 0 \label{newconstraint2}
\end{eqnarray}
From (\ref{newconstraint1}) we get 12 constraints while (\ref{newconstraint2}) gives 18 because of the antisymmetry in $IJ$.
\end{itemize}

The above constraints cannot all be first class. Also note that the constraints (\ref{newconstraint1}) and (\ref{newconstraint2}) are different
from the second-class constraints considered in the Hamiltonian analysis of Hilbert-Palatini action \cite{sa,peldan}.

Before calculating the constraint algebra we note another interesting feature in our theory. While the Gauss constraint
$\mathcal{G}_{IJ}$ generates the $SO(3,1)$ transformations, the constraint which actually generates the spatial diffeomorphisms is a combination
given by
\begin{eqnarray}
\tilde{H_a}:= H_a +  \omega_a^{~IJ} \mathcal{G}_{IJ} +
\frac{1}{\alpha} \epsilon_{abc} C^b_I \Pi^c_I \label{spatialdiffeo}
\end{eqnarray}
This can be easily demonstrated as:
\begin{eqnarray}
\delta^{\tilde{H_a}} \omega_c^{~IJ}  &:=& \left\{ \omega_c^{~IJ},\tilde{H_a}(N^a)\right\} =
N^a \partial_a \omega_c^{~IJ} + \omega_a^{~IJ} \partial_c N^a = \mathcal{L}_{N^a}\omega_c^{~IJ} ,\nonumber \\
\delta^{\tilde{H_a}} \Pi^c_{IJ} &:=& \left\{ \Pi^c_{IJ}, \tilde{H_a}(N^a)\right\} =
N^a \partial_a \Pi^c_{IJ} - \Pi^a_{IJ} \partial_a N^c +  \Pi^c_{IJ} \partial_a N^a   = \mathcal{L}_{N^a}\Pi^c_{IJ} ,\nonumber \\
\delta^{\tilde{H_a}} V_c^I &:=&  \left\{  V_c^I,\tilde{H_a}(N^a)\right\} =
N^a \partial_a  V_c^I +  V_a^I \partial_c N^a = \mathcal{L}_{N^a} V_c^I  ,\nonumber \\
\delta^{\tilde{H_a}}\Pi^c_I &:=&  \left\{ \Pi^c_I, \tilde{H_a}(N^a)\right\} =
N^a \partial_a \Pi^c_I  - \Pi^a_I  \partial_a N^c + \Pi^c_I \partial_a N^a = \mathcal{L}_{N^a}\Pi^c_{I}
\end{eqnarray}
The geometrical meaning of $\tilde{H_a}$ makes it easy to calculate its Poisson brackets with other constraints.
Including all the primary constraints found so far, we can write the total Hamiltonian as
\begin{eqnarray}
H_T := N H + N^a \tilde{H_a} + \Lambda^{IJ} \mathcal{G}_{IJ} +
\gamma_a^I C^a_I + \lambda_a^{IJ} \Phi^a_{IJ} \label{htotal1}
\end{eqnarray}
where the expressions of the constraints are given by equations (\ref{hamiltonian1}),(\ref{spatialdiffeo}),(\ref{gauss1}),(\ref{newconstraint1}) and
(\ref{newconstraint2}) respectively.

We now calculate the algebra generated by the constraints. It turns out that the terms which are not weakly zero are:
\begin{eqnarray}
\left\{H(N), \Phi^a_{IJ}(\lambda_a^{IJ}) \right\} &=&
- N N_I \Pi^a_J \left(\lambda_a^{IJ} +\frac{1}{\alpha}\epsilon_{IJKL}\lambda_a^{KL}\right) ,\label{constalg1}\\
\left\{ C^a_I(\gamma_a^I),\Phi^b_{JK}(\lambda_b^{JK}) \right\} &=&
\epsilon^{abc} \gamma_b^I V_c^J\left(\alpha \lambda_a^{IJ} +\epsilon_{IJKL}\lambda_a^{KL}\right) \label{constalg2}
\end{eqnarray}
At this point we can introduce further secondary constraints or try
to solve for some of the Lagrangian multipliers. Before that, let us
again consider the degrees of freedom. Clearly $\mathcal{G}_{IJ}$
and $\tilde{H_a}$ are first class. Since the Hamiltonian formulation
should be equivalent to the Lagrangian one for consistency, we would
like $H$ to also be first class although it is not at this moment. All together we would have $(6+3+1)=10$
first-class constraints removing $20$ degrees of freedom. The
constraints $C^a_I$ and $\Phi^a_{IJ}$ are second class removing
$12+18=30$ degrees of freedom. To obtain the torsion-free case, we
need to find $6$ more constraints hidden in the Eqs
(\ref{constalg1}) and (\ref{constalg2}). Note that for a consistent
Hamiltonian system we require
\begin{eqnarray}
\dot{\Phi}^a_{IJ}(\sigma_a^{IJ}) := \left\{\Phi^a_{IJ}(\sigma_a^{IJ}),H_T \right\} &\approx& 0  ,  \label{phievol1} \\
\dot{C^a_I}(\eta_a^I):= \left\{C^a_I(\eta_a^I),H_T \right\} &\approx& 0 \label{Cevol1}
\end{eqnarray}
for arbitrary smearing functions $\sigma_a^{IJ}$ and $\eta_a^I$. Explicitly evaluating (\ref{phievol1}) using Eqs (\ref{constalg1}) and (\ref{constalg2}) we get
\begin{eqnarray}
\dot{\Phi}^a_{IJ} \approx 0 ~~ &\Rightarrow& -\epsilon^{abc} \gamma_{b[I} V_{J]c} + \frac{1}{\alpha}N N_{[I} \Pi^a_{J]} \approx 0 \label{phievol2}\\
&\Rightarrow& \left(\gamma_b^I V_c^J -\gamma_c^I V_b^J - \gamma_b^J V_c^I +\gamma_c^J V_b^I  \right) -
\epsilon_{abc}\frac{N}{\alpha}\left(N^I \Pi^{a J} - N^J \Pi^{a I}  \right) \approx 0 \label{phievol3}
\end{eqnarray}
Multiplying (\ref{phievol3}) with $V^b_J$ and using the properties (\ref{parameterconstraints2}) we get
\begin{eqnarray}
2 \gamma_c^I +V^b_J\gamma_b^J V_c^I - V^b_J \gamma_c^J V_b^I + \epsilon_{abc}\frac{N}{\alpha} N^I \Pi^{a J}V^b_J \approx 0 \label{phievol4}
\end{eqnarray}
By multiplying this equation with $N_I$, $V^c_I$ and $V_d^I$ respectively and using the relations (\ref{parameterconstraints1}) and
(\ref{parameterconstraints2}) we obtain the following relations
\begin{eqnarray}
\gamma_c^I N_I &=& \frac{N}{2\alpha} \epsilon_{abc} \Pi^a_J V^b_J ,\label{phievolrel1}\\
\gamma_c^I V^c_I &=& 0 \label{phievolrel2} ,\\
\gamma_c^I V_d^I &=& 0 \label{phievolrel3}
\end{eqnarray}
where we have used Eq.(\ref{phievolrel2}) to obtain (\ref{phievolrel3}).
Finally from the equations (\ref{phievolrel1}) and (\ref{phievolrel3}) we get a solution for the Lagrangian multiplier $\gamma_c^I$ as
\begin{eqnarray}
\gamma_c^I &=& - \frac{N}{2\alpha} \epsilon_{abc} N^I \Pi^a_J V^b_J \label{gammasoln}
\end{eqnarray}
So, we have obtained 12 components of $\gamma_a^I$ from the 18 equations in (\ref{phievol2}). Consequently there are 6 constraints remaining. By
inserting the solutions (\ref{gammasoln}) back into (\ref{phievol2}) we get the following constraint:
\begin{eqnarray}
\Pi^a_J + \Pi^a_K N^K N^J + \Pi^b_K V^{a K} V_b^J \approx 0 \nonumber
\end{eqnarray}
It can be shown that this constraint is actually equivalent to
\begin{eqnarray}
\Pi^a_{I} V^{b}_I + \Pi^b_{I} V^a_I \approx 0 \label{proofpi01}
\end{eqnarray}
Rewriting this in terms of basic variables we get the desired 6 secondary constraints
\begin{eqnarray}
\chi^{ab} := -\frac{1}{\sqrt q}\left(\Pi^{(a}_I \Pi^{b)IJ} N_J \right) \approx 0 \label{newconstraint3}
\end{eqnarray}
In the above analysis we have used the relations (\ref{parameterconstraints1}) and (\ref{parameterconstraints2}).

Note that something similar occurs in various analysis performed in PGT \cite{nikolic}. However, we have managed to explicitly extract all the
second-class constraints. Of course, our constraint analysis is not over because we have to ensure that the enlarged constraint algebra is now closed.
We cannot add further secondary constraints to close the algebra but have to consistently solve for the Lagrangian multipliers. It turns out that
$\chi^{ab}$ have non-zero Poisson brackets with all constraints except for itself, Gauss and  spatial diffeomorphism constraint. They are calculated
as
\begin{eqnarray}
\left\{\chi^{ab}(\sigma_{ab}),H(N)\right\} &=&
\frac{\sigma_{ac}}{\sqrt q}\Pi^a_{[I}N_{J]}\left[
- D_b\left(\frac{N}{\sqrt q}\Pi^b_{IL}\Pi^c_{JL}\right) +  D_b\left(\frac{N}{\sqrt q}\Pi^b_{JL}\Pi^c_{IL}\right) -\frac{1}{2}N N_{[I}\Pi^c_{J]}
\right]\nonumber\\
&& + \frac{2\sigma_{ab}}{\sqrt q} \Pi^a_K N_I V^c_J \Pi^b_{KJ}D_c(N N^I) -  \frac{2\sigma_{cb}}{ q} N N^J \Pi^b_{IJ}\Pi^c_{IK} D_a \Pi^a_K  ,
\label{constalg3}\\
\left\{\chi^{ab}(\sigma_{ab}),C^c_I(\gamma_c^I)\right\} &=&
\frac{\alpha \sigma_{ac}}{2 \sqrt q}\Pi^a_{[I}N_{J]}\epsilon^{cdb}\gamma_d^{[I}V_b^{J]} +
\frac{2\sigma_{ab}}{\sqrt q}\Pi^a_K N_I V^c_J \Pi^b_{KJ}\gamma_c^I - \frac{2\alpha \sigma_{cb}}{\sqrt q}N_J \Pi^b_{IJ}\epsilon^{cad} D_a \gamma_d^I
,\label{constalg4}\nonumber\\
&&\\
\left\{\chi^{ab}(\sigma_{ab}),\Phi^c_{IJ}(\lambda_c^{IJ})\right\} &=&
\frac{2\sigma_{cb}}{\sqrt q}N_J \Pi^b_{IJ}\epsilon^{cad}\epsilon_{KLIM} V_d^M \lambda_a^{KL} \label{constalg5}
\end{eqnarray}
These along with (\ref{constalg1}) and (\ref{constalg2}) are the non-zero terms in the constraint algebra.

At this point we show a result which will be used a number of times subsequently to show the closure of the constraint algebra.
Let $\rho^{KL}$ be a function antisymmetric in the indices $KL$. We suppress the spatial indices as they are not important for this calculation.
Suppose for any real number $r$ we have
\begin{eqnarray}
\rho^{MN} \epsilon_{MNIJ} + r \rho^{IJ} = 0  \label{lagmulproof1}
\end{eqnarray}
It is easy to see that this implies
\begin{eqnarray}
&&\rho^{IJ}\epsilon_{KLIJ} + \frac{1}{r} \rho^{MN} \epsilon_{MNIJ}\epsilon_{KLIJ} = 0 \nonumber\\
\mbox{or,}~~~&& -r \rho^{KL} -\frac{2}{r}\rho^{MN}\delta^{[K}_M \delta^{L]}_N = 0\nonumber \\
\Rightarrow~~&& \left(r + \frac{4}{r} \right) \rho^{IJ} = 0 ~~~\Rightarrow~~ \rho^{IJ} = 0 \nonumber
\end{eqnarray}
Consequently we get the following result
\begin{eqnarray}
\rho^{MN} \epsilon_{MNIJ} + r \rho^{IJ} = 0 ~~ \Rightarrow~~ \rho^{IJ} = 0 \label{lagmulproof2}
\end{eqnarray}
where we have used Eq.(\ref{lagmulproof1}) and the properties of $SO(3,1)$ Levi-Civita symbols.

Also, it is possible to use the constraints $C^a_I$ and $\Phi^a_{IJ}$ to derive a relation:
\begin{eqnarray}
D_{IJ} := D_a \Pi^a_{IJ} -  \frac{1}{\alpha}\epsilon_{IJKL} \Pi^{a K} V_a^L \approx 0 \label{defDIJ}
\end{eqnarray}
From the Gauss constraint (\ref{gauss1}) and equation (\ref{defDIJ}) and the result of (\ref{lagmulproof2}) we get
\begin{eqnarray}
\Pi^c_{I} V_{J c} - \Pi^c_{J} V_{I c} \approx 0 \label{proofpi02}
\end{eqnarray}
Multiplying Eq.(\ref{proofpi02}) with (a) $N_J$ and then $V^b_I$ and (b)$V^b_I$ and then $V^a_J$, we get respectively
\begin{eqnarray}
\Pi^b_{I} N^I &\approx& 0, \label{proofpi03}\\
\Pi^a_{I} V^{b}_I - \Pi^b_{I} V^a_I &\approx& 0 \label{proofpi04}
\end{eqnarray}
where we have again made use of the relations (\ref{parameterconstraints1}) and (\ref{parameterconstraints2}).
From equations (\ref{proofpi01}) and (\ref{proofpi04}) we obtain $\Pi^a_{I} V^{b}_I \approx 0$. Multiplying this with $V_b^J$ and
using (\ref{proofpi03}) and (\ref{parameterconstraints2}), we get $\Pi^a_I \approx 0$. So the constraints
$\mathcal{G}_{IJ},C^a_I,\Phi^a_{IJ},\chi^{ab}$ together imply torsion free condition.

As a consequence, the constraint algebra is extremely simplified. It is easy to see that the Hamiltonian constraint now becomes first class because the
brackets (\ref{constalg1}) and (\ref{constalg3}) are now weakly zero. Thus our previous wish can be
fulfilled.
The other three constraints remain second class. But the non-zero terms
become much more simple as
\begin{eqnarray}
\left\{ C^a_I(\gamma_a^I),\Phi^b_{JK}(\lambda_b^{JK}) \right\} &=&
\epsilon^{abc} \gamma_b^I V_c^J\left(\alpha \lambda_a^{IJ} +\epsilon_{IJKL}\lambda_a^{KL}\right) ,\label{newconstalg1}
\\
\left\{\chi^{ab}(\sigma_{ab}),C^c_I(\gamma_c^I)\right\} &=&
- \frac{2\alpha \sigma_{cb}}{\sqrt q}N^J \Pi^b_{IJ}\epsilon^{cad} D_a \gamma_d^I ,\label{newconstalg2}
\\
\left\{\chi^{ab}(\sigma_{ab}),\Phi^c_{IJ}(\lambda_c^{IJ})\right\} &=&
\frac{2\sigma_{cb}}{\sqrt q}N_J \Pi^b_{IJ}\epsilon^{cad}\epsilon_{KLIM} V_d^M \lambda_a^{KL}\label{newconstalg3}
\end{eqnarray}
It is not difficult to check that except for Eqs (\ref{newconstalg1}-\ref{newconstalg3}) the constraint algebra is weakly closed.
We can easily solve the equations $\dot{C^a_I} \approx 0$, $\dot{\chi^{ab}} \approx 0$ and $\dot{\Phi}^c_{JK} \approx 0$ with arbitrary smearing
functions and get the solutions of the Lagrange multipliers $\lambda_c^{JK}=0 $ and $\gamma_a^I \approx 0$. Thus we go back to Hamiltonian
formulation of general relativity without torsion which has been well studied \cite{peldan}.

\section{Concluding remarks} \label{sec4}

In general relativity, one usually works with torsion-free connections, and thus gravitational degrees of freedom are encoded only in the metric tensor. However, in
general, from the perspective of local Poincare invariance,  it is possible to introduce further gravitational degrees of freedom via an
independent torsion tensor. Furthermore, there are models where torsion can be used to explain the current acceleration of our universe\cite{nester}.
However, a well-defined dynamical theory requires a consistent Hamiltonian description. Moreover, to incorporate torsion into a theory of quantum gravity
built from the canonical perspective, it is essential to have a well-defined Hamiltonian formalism.  Our goal in this paper is to obtain a consistent
Hamiltonian analysis via the Dirac procedure of a typical action containing torsion.

Our action contains a term quadratic in torsion apart from the
Hilbert-Palatini term. In fact, this torsion term is just the
difference between the Nieh-Yan term and the Holst term. Although
the final result is the expected torsion-free case, we learnt
several interesting lessons in our analysis, which may be important
for further studies of torsion in the Hamiltonian framework. The key
points of our analysis may be summarized as follows.
\begin{enumerate}[(i)]
\item We construct the spatial diffeomorphism constraint $\tilde{H_a}$ in equation (\ref{spatialdiffeo}). Unlike the torsion-free case, $\tilde{H_a}$
contains an additional term $\frac{1}{\alpha}\epsilon_{abc} C^b_I \Pi^c_I$ because of the torsion term in action (\ref{action}). Without adding this
term we cannot ensure that the diffeomorphism constraint generates Lie derivatives of the basic variables.
\item When we get six new second-class constraints (\ref{newconstraint3}) from the consistency condition of constraint $\Phi^a_{IJ}$, we do two things
at the same time. One is to solve for Lagrangian multiplier $\gamma_c^I$, while the other is to determine secondary constraint $\chi^{ab} $.
They are related to each other. This is certainly different from the analysis in Hilbert-Palatini theory \cite{peldan} as well as the theory studied
in \cite{shyam}. We also manage to get an explicit expression of this new set of secondary second-class constraints which has not been done before.
\item We get $\Pi^a_I \approx 0$ from the constraints $\mathcal{G}_{IJ},C^a_I,\Phi^a_{IJ}$ and $\chi^{ab}$ and prove that the constraint algebra is
closed. The procedure of the proof is very interesting and may find application in other Hamiltonian systems.
\item We do not fix any gauge before performing the Hamiltonian analysis, because gauge fixing in systems with second-class constraints might give rise
to inconsistencies and some equations of motion of original theory might be lost.
\end{enumerate}

In this paper we demonstrate a procedure to obtain a consistent Hamiltonian analysis of the action with torsion. For future investigation, we may consider
matter coupling in this action. For example, in the case there is Dirac field, torsion is non-vanishing. It will be
interesting to analyze such an action following our procedure. We may also use the results of this paper to analyze actions like $R+R^{2}$ with non-zero torsion. Torsion is dynamical in such models even without matter coupling. Studying models with dynamical torsion by Hamiltonian analysis will
give us a deeper understanding of the dynamical behaviour related to torsion. This may further shed light on how to incorporate torsion into theories
of quantum gravity.\\

{\bf Acknowledgements:}\\

This work is supported in part by  NSFC (Grant No. 10975017) and  the Fundamental Research Funds for the Central Universities.
JY would like to acknowledge the support of NSFC (Grant No. 10875018). KB would also like to thank China Postdoctoral Science Foundation
(Grant No.20100480223) for financial support.

\appendix
\section{Proof of Local Poincare Invariance}

In this section, we  prove that the action (\ref{action}) is invariant under local Poincare transformations. We shall prove this for the two terms,
$eR$ and $\epsilon^{\mu\nu\alpha\beta}T^{I}_{~\mu\nu}T_{I\alpha\beta} $ separately.

At first we prove local translational symmetry of the two terms. The determinant of co-tetrad reads
\begin{equation}
e=\frac{1}{4!}\epsilon_{IJKL}\epsilon^{\mu\nu\alpha\beta}e^{I}_{\mu}e^{J}_{\nu}e^{K}_{\alpha}e^{L}_{\beta} \nonumber
\end{equation}
Under infinitesimal local translational transformation, the co-tetrad, tetrad and spin connection transform respectively as \cite{blagojevic3lectures}
\begin{eqnarray}
\delta_{0}
e^{K}_{\mu}&=&-e^{K}_{\lambda}\partial_{\mu}\xi^{\lambda}-\xi^{\lambda}\partial_{\lambda}e^{K}_{\mu} ,\nonumber\\
\delta_{0}e^{\mu}_{K}&=&e^{\lambda}_{K}\partial_{\lambda}\xi^{\mu}-\xi^{\lambda}\partial_{\lambda}e^{\mu}_{K},\nonumber\\
\delta_{0}\omega_{\mu}^{IJ}&=&-\omega_{\lambda}^{IJ}\partial_{\mu}\xi^{\lambda}-\xi^{\lambda}\partial_{\lambda}\omega_{\mu}^{IJ}\nonumber
\end{eqnarray}
where the gauge group parameters $\xi^{\mu}$ are functions of space-time points.

Up to first order we have \cite{MM}
\begin{eqnarray}
e'&=&e-\xi^{\lambda}\partial_{\lambda}e-e\partial_{\lambda}\xi^{\lambda},\\
R'_{\mu\nu}\,^{KL}&=& R_{\mu\nu}\,^{KL}-\xi^{\lambda}\partial_{\lambda}R_{\mu\nu}\,^{KL}
-R_{\mu\lambda}\,^{KL}\partial_{\nu}\xi^{\lambda}+R_{\nu\lambda}\,^{KL}\partial_{\mu}\xi^{\lambda}, \\
T'^{I}_{\mu\nu}&=& T^{I}_{\mu\nu}-\xi^{\lambda}\partial_{\lambda}T^{I}_{\mu\nu}
-T^{I}_{\lambda\nu}\partial_{\mu}\xi^{\lambda}+T^{I}_{\lambda\mu}\partial_{\nu}\xi^{\lambda}
\end{eqnarray}
Under above transformations, $R=e^{\mu}_{I}e^{\nu}_{J}R_{\mu\nu}\,^{IJ}$ transforms to
\begin{equation}
R'=e'^{\mu}_{I}e'^{\nu}_{J}R'_{\mu\nu}\,^{IJ}=R-\xi^{\lambda}\partial_{\lambda}R.
\end{equation}
Therefore for the Hilbert-Palatini term we get
\begin{equation}
e'R'=eR-\partial_{\lambda}(\xi^{\lambda}eR). \label{3}
\end{equation}
where the additional term is just a total divergence which can be ignored.

For the $S_T$ term we have
\begin{eqnarray}
\epsilon^{\mu\nu\alpha\beta}T'^{I}_{~\mu\nu}T'_{I\alpha\beta} &=& \epsilon^{\mu\nu\alpha\beta}
\left(T^{I}_{~\mu\nu}-\xi^{\lambda}\partial_{\lambda}T^{I}_{~\mu\nu}
-T^{I}_{~\lambda\nu}\partial_{\mu}\xi^{\lambda}+T^{I}_{~\lambda\mu}\partial_{\nu}\xi^{\lambda}\right)\nonumber\\
&&\hspace{6em}\left(T_{I\alpha\beta}-\xi^{\sigma}\partial_{\sigma}T_{I\alpha\beta}
-T_{I\sigma\beta}\partial_{\alpha}\xi^{\sigma}+T_{I\sigma\alpha}\partial_{\beta}\xi^{\sigma}\right)\nonumber\\
& =&
\epsilon^{\mu\nu\alpha\beta}T^{I}_{~\mu\nu}T_{I\alpha\beta}+\epsilon^{\mu\nu\alpha\beta}T^{I}_{~\mu\nu}T_{I\alpha\beta}\partial_{\lambda}\xi^{\lambda}
-4\epsilon^{\mu\nu\alpha\beta}T^{I}_{~\mu\nu}T_{I\lambda\beta}\partial_{\alpha}\xi^{\lambda}.
\label{5}
\end{eqnarray}
Since
\begin{eqnarray}
T_{I\lambda\beta}=\frac{1}{2}T_{I\rho\sigma}\delta^{[\rho}_{\lambda}\delta^{\sigma]}_{\beta}=-\frac{1}{4}T_{I\rho\sigma}
\epsilon^{\rho\sigma\gamma\delta}\epsilon_{\lambda\beta\gamma\delta}
\end{eqnarray}
we may rewrite the last term of (\ref{5}) as
\begin{eqnarray*}
4\epsilon^{\mu\nu\alpha\beta}T^{I}_{~\mu\nu}T_{I\lambda\beta}\partial_{\alpha}\xi^{\lambda}&=&-
\epsilon^{\mu\nu\alpha\beta}\epsilon_{\lambda\gamma\delta\beta}\epsilon^{\rho\sigma\gamma\delta}T^{I}_{~\mu\nu}T_{I\rho\sigma}\partial_{\alpha}\xi^{\lambda}
\\
&=&\delta^{[\mu}_{\lambda}\delta^{\nu}_{\gamma}\delta^{\alpha]}_{\delta}\epsilon^{\rho\sigma\gamma\delta}
T^{I}_{~\mu\nu}T_{I\rho\sigma}\partial_{\alpha}\xi^{\lambda}\\
&=&-4\epsilon^{\mu\nu\alpha\beta}T^{I}_{~\mu\nu}T_{I\lambda\beta}\partial_{\alpha}\xi^{\lambda}
+2\epsilon^{\mu\nu\rho\sigma}T^{I}_{~\mu\nu}T_{I\rho\sigma}\partial_{\lambda}\xi^{\lambda}
\end{eqnarray*}
which implies that
\begin{eqnarray}
\epsilon^{\mu\nu\alpha\beta}T^{I}_{~\mu\nu}T_{I\alpha\beta}\partial_{\lambda}\xi^{\lambda}
=4\epsilon^{\mu\nu\alpha\beta}T^{I}_{~\mu\nu}T_{I\lambda\beta}\partial_{\alpha}\xi^{\lambda}
\end{eqnarray}
Putting this back into (\ref{5}) we see that the last two terms of (\ref{5}) vanish thereby giving the desired result.
Hence, we can say $eR=ee^{\mu}_{I}e^{\nu}_{J}R_{\mu\nu}\,^{IJ}$ and $\epsilon^{\mu\nu\alpha\beta}T^{I}_{~\mu\nu}T_{I\alpha\beta}$ are
local translational invariant.

Next, we give proof on local Lorentz symmetry. Under infinitesimal local Lorentz transformations, the co-tetrad, tetrad and spin
connection transform as
\begin{eqnarray}
\delta_{0} e^{K}_{\mu}&=&A^{K}{_S}e^{S}_{\mu} ,\label{rotation} \\
\delta_{0} e^{\mu}_{K}&=&A_{K}{^S}e^{\mu}_{S} ,\\
\delta_{0}\omega_{\mu}^{IJ}&=&A^{I}{_K}\omega_{\mu}^{KJ}+A^{J}{_K}\omega_{\mu}^{IK}-\partial_{\mu}A^{IJ}
\end{eqnarray}
where gauge group parameters $A^{IJ}$ are also functions of space-time points, and $A^{IJ}=-A^{JI}$. It is obvious that $e$ is
invariant under rotation (\ref{rotation}), while the curvature transforms to \cite{MM}
\begin{equation}
R'_{\mu\nu}\,^{IJ}=R_{\mu\nu}\,^{IJ}+A^{I}{_M}R_{\mu\nu}\,^{MJ}+A^{J}{_M}R_{\mu\nu}\,^{IM}
\end{equation}
The term $eR=ee^{\mu}_{I}e^{\nu}_{J}R_{\mu\nu}\,^{IJ}$ has local Lorentz symmetry since
\begin{equation}
R'=e'^{\mu}_{I}e'^{\nu}_{J}R'_{\mu\nu}\,^{IJ}=R
\end{equation}
while the torsion transforms to \cite{MM}
\begin{equation}
T'^{I}_{~\mu\nu}=T^{I}_{~\mu\nu}+A^{I}{_J}T^{J}_{\mu\nu} \label{a15}
\end{equation}
Using Eq.(\ref{a15}) we can obtain the invariance of the $T^2$ term under local Lorentz rotations as
\begin{eqnarray}
\epsilon'^{\mu\nu\alpha\beta}T'^{I}_{~\mu\nu}T'_{I\alpha\beta}&=&\epsilon^{\mu\nu\alpha\beta}(T^{I}_{~\mu\nu}+A^{I}{_J}T^{J}_{\mu\nu})
(T_{I\alpha\beta}+A_{I}{^K}T_{K\alpha\beta}) \nonumber\\
&=&\epsilon^{\mu\nu\alpha\beta}T^{I}_{~\mu\nu}T_{I\alpha\beta},
\end{eqnarray}

In summary, the action (\ref{action}) is invariant under local Poincare transformation, which gives a well-defined Poincare gauge theory of gravity.


\end{document}